\documentclass[runningheads]{svmult}

\usepackage{makeidx}   
\usepackage{graphicx}  
\usepackage{subeqnar}  
\usepackage{multicol}  
\usepackage{physprbb}  
\makeindex             

\newcommand{\greeksym}[1]{{\usefont{U}{psy}{m}{n}#1}}
\newcommand{\umu}{\mbox{\greeksym{m}}}

\begin{document}
\title*{Correcting the chromatic and airmass dependent extinction for
        TIMMI2 spectra}
\toctitle{Correcting the chromatic and airmass dependent extinction for
          TIMMI2 spectra}
%
%
\titlerunning{Correcting the chromatic and airmass dependent extinction}
%
\author{Oliver Sch\"utz\inst{1}
\and Michael Sterzik\inst{2}}
\authorrunning{Sch\"utz \& Sterzik}
%
%
\institute{Max-Planck-Institut f\"ur Astronomie, K\"onigstuhl 17,
           69117 Heidelberg, Germany
\and European Southern Observatory, Alonso de Cordova 3107,
     Santiago~19, Chile}

\maketitle              

\begin{abstract}
We present a method to correct the chromatic and airmass dependent
extinction for N-band spectra taken with the TIMMI2 instrument at the
ESO\,/\,La~Silla observatory. Usually, the target and calibrator star
have to be observed at similar airmass in order to obtain reliable
spectrophotometric fluxes. Our method allows to correct the atmospheric
extinction and substantially improves the spectrophotometric flux
calibration, when the standard stars were observed at a very different
airmass than the targets. Hundreds of standard star measurements in
several passbands (N1, N8.9, N10.4, N11.9) were used to derive mid-IR
extinction coefficients. We demonstrate that applying our correction
of the differential extinction to test data results in a
spectrophotometric accuracy up to 2\% within the literature flux.
\end{abstract}

\vspace{0cm}

\section{The influence of atmospheric extinction in the mid-IR}
For spectroscopy the target stars and calibrators should generally be
observed at a comparable airmass to avoid a varying extinction between
the two objects. Different to near-IR observations, in the mid-IR a
good calibrator at the same airmass is often difficult to find and the 
theoretical behaviour of the extinction with airmass and wavelength is
not really known up to now. Some authors claim there would be no clear 
dependence and thus for photometry no airmass correction needs to be 
applied. However these authors used dozens of standard star observations, 
while we apply several hundred measurements.

In mid-IR spectroscopy the influence of extinction is more evident
than for photometry. No good spectral flux calibrations are possible
unless the target and calibrator star are observed close in time and
very close in airmass ($\leq$~0.1 airmass distance). Therefore, some
observers often use additional photometric measurements to correct the
spectral fluxes. Since the extinction as a function of wavelength varies
significantly within the N-band, also the spectral slope needs to be
corrected, if the target and calibrator had been taken at different
airmass. 

Our goal is to explain the extinction in the mid-IR as a function of
airmass. We further demonstrate that the extinction has a non-linear
wavelength dependence within the N-band.

\section{Data analysis}
All observations of standard stars obtained with 
TIMMI2\footnote{http://www.ls.eso.org/lasilla/sciops/timmi}
are archived from the beginning of operations in 2001 until present. 
The conversion factors between measured counts and known fluxes depend 
on the filter and lens scale used as well as the airmass (all 
systematically), but also vary with the sky conditions (statistically). 
A decreasing count rate as a function of rising airmass corresponds to 
an increasing conversion factor towards larger airmass. 

We first separate all standard star measurements according to the filter 
and lens scale used. Data obtained before the TIMMI2 upgrade in 
October 2002 are treated separately because of different electronical 
gains in the previous readout system, but after a normalisation with 
the conversion factor for airmass AM~= 1.0 both results are comparable. 

In Fig.\,1 we show the conversion factors (hereafter called Conv) for
the N1 filter as a function of airmass. A more detailed version of this 
article with further illustrations will be available via the TIMMI2 
webpage (cf. footnote). Note that a clear trend for Conv increasing 
with airmass can be seen, especially if you look towards the lower 
margin of the distribution. 

{\bf Why can we constrain our analyse to the lower margin\,?} For a
given detector and instrument configuration there exists a well
defined optimal sensitivity for a star, i.e. the least extincted count
rate achieved for best weather conditions. This corresponds to a
minimum conversion factor. On the other side, the sensitivity 
deteriorates arbitrarily with worse sky conditions, there is no clear 
maximum for the conversion factor. This is the reason why the 
distribution of points in Fig.\,1 seems to be confined in lower 
y-direction but spreads towards higher y.

\begin{figure}[t]
  \begin{center}
    \includegraphics[width=.67\textwidth]{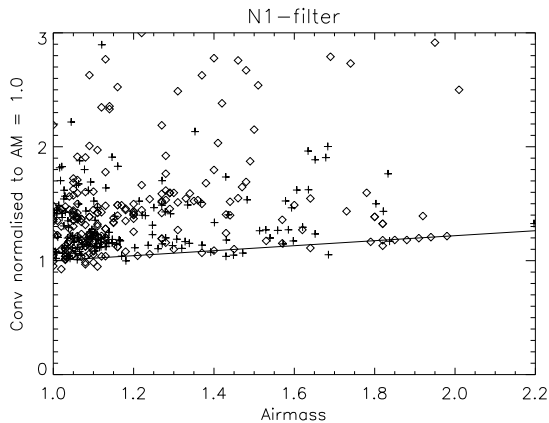}
  \end{center}
  \caption{Conversion factors as a function of airmass are displayed
           for the N1 passband. Squares are standard star measurements 
           obtained since the TIMMI2 upgrade in October 2002, crosses 
           symbolise data taken before this time. Especially watch the
           well defined rising of points between AM = 1.8 and AM = 2. 
           The solid line represents the fit from which formula~(1) is
           obtained}
\end{figure}

Since we need measurements obtained under identical weather conditions,
it is sufficient to make a fit only with the lower-value points. We 
normalise for each filter the conversion factors by the value for AM
= 1.0, in order to make comparable the measurements obtained before and
after the TIMMI2 upgrade as well as to make these results comparable 
to other instruments.

\section{Results}

\subsection{Differential extinction as function of airmass}

We present relations, deduced as explained in Sect.\,2, which describe
the dependence of the atmospheric extinction with airmass (AM). 
Equation~(1a), for example, signifies that at this wavelength the
(spectral) flux must be corrected by 22\% to account for the
extinction between AM = 1.0 and AM = 2.0. For mid-IR filters not given
below the number of standard star measurements was not sufficient to 
calculate a firm result. Since the TIMMI2 archival of standard stars
is an ongoing effort, we can add results for these passbands at a
later stage. 

In (1a--1d) AM represents the airmass of observation and Corr is the
factor to correct the targets' flux into measurements at airmass 1.0. 
Note that the TIMMI2 filters are not always named according to their
central wavelength $\lambda_0$.

\begin{subeqnarray}
\mathrm{N1}    \enspace (\lambda_0 = 8.6  \, \umu\mathrm{m}): \quad
      \mathrm{Corr(AM) = 1 + 0.220 * (AM-1)} \\
\mathrm{N8.9}  \enspace (\lambda_0 = 8.7  \, \umu\mathrm{m}): \quad
      \mathrm{Corr(AM) = 1 + 0.208 * (AM-1)} \\
\mathrm{N10.4} \enspace (\lambda_0 = 10.3 \, \umu\mathrm{m}): \quad
      \mathrm{Corr(AM) = 1 + 0.212 * (AM-1)} \\
\mathrm{N11.9} \enspace (\lambda_0 = 11.6 \, \umu\mathrm{m}): \quad
      \mathrm{Corr(AM) = 1 + 0.116 * (AM-1)}
\end{subeqnarray}

For spectroscopy, both the target and calibrator must be flux
corrected before their division. The wavelength dependence of Corr has
to be taken into account. In a first approach, we use a linear
interpolation between Corr for N1 and N11.9, since the best fits are
achieved for those filters:

\begin{equation}
\mathrm{Corr}(\lambda, \mathrm{AM}) 
= 1 + [0.220 - (0.220-0.116) / 3 * (\lambda - 8.6 \, \umu\mathrm{m})] 
  * (\mathrm{AM}-1)
\end{equation}

The factor '3' in (2) expresses the difference in wavelength between
N1 and N11.9. Finally, $\mathrm{Corr}(\lambda, \mathrm{AM})$ is 
multiplied with the uncorrected spectral flux $F(\lambda, \mathrm{AM})$:

\begin{equation}
F_{\mathrm{real}}\,(\lambda) 
= F(\lambda, \mathrm{AM}) * \mathrm{Corr}(\lambda, \mathrm{AM})
\end{equation}

In any data reduction script which also extracts the airmass from the
file headers, (2) and (3) can be conveniently included. With this 
approximation already a significant improvement of the spectral fluxes 
is seen in Fig.\,2, especially for large distances in airmass. At a
later stage, when we can include further correction factors for other 
N-band filters, the flux correction can be improved to an even higher 
perfection.

\subsection{N-band extinction coefficients}

From the relative increase of atmospheric extinction between AM~= 1.0
and AM = 2.0 we can, in principle, calculate the extinction
coefficients $K$ for La~Silla (usually given in mag\,/\,AM). These
values depend on the observatory site, especially on the
altitude and climatic conditions. 

Table~1 summarises the median N-band extinction coefficients. Thereof 
an unextincted photometry with magnitudes $m_{\mathrm{real}}$ can be
obtained via the relation

\begin{equation}
m_{\mathrm{obs}}(\mathrm{AM}) = m_{\mathrm{real}} + K * \mathrm{AM} \, .
\end{equation}

\begin{table}[t]
\caption{Median N-band extinction coefficients for La Silla}
  \begin{center}
    \setlength\tabcolsep{15pt}
    \begin{tabular}{lc}
      \hline\noalign{\smallskip}
      Filter & $K$\,[mag/AM] \\
      \noalign{\smallskip}
      \hline
      \noalign{\smallskip}
      N1     &  0.22         \\
      N8.9   &  0.21         \\
      N10.4  &  0.21         \\
      N11.9  &  0.12         \\
      \hline
    \end{tabular}
  \end{center}
\end{table}

The wavelength dependence of our extinction coefficients is similar to
data for Mauna Kea, while the absolute values differ because of the
other altitude.

\section{Application to spectroscopic data}

Formula (2) and (3) were tested with a 
cross-calibration of different standard stars taken from the same night. 
We used test data obtained both in December 2002 and September 2003. A
calibrator at low airmass is taken while the objects were observed at
higher airmass (see the headers of Fig.\,2 for the actual airmass
value).

\begin{figure}[t]
  \begin{center}
    \includegraphics[width=.495\textwidth]{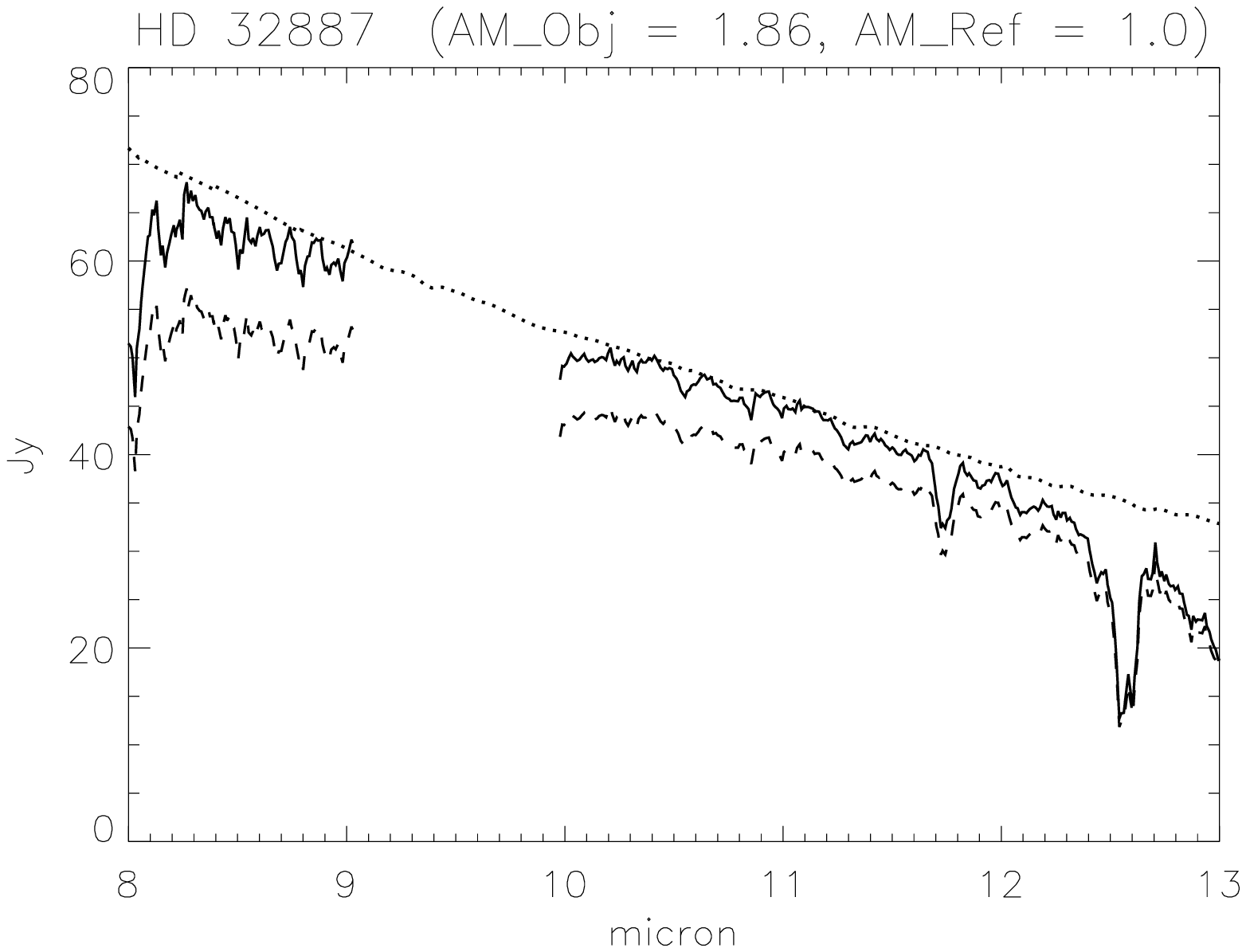}
    \includegraphics[width=.495\textwidth]{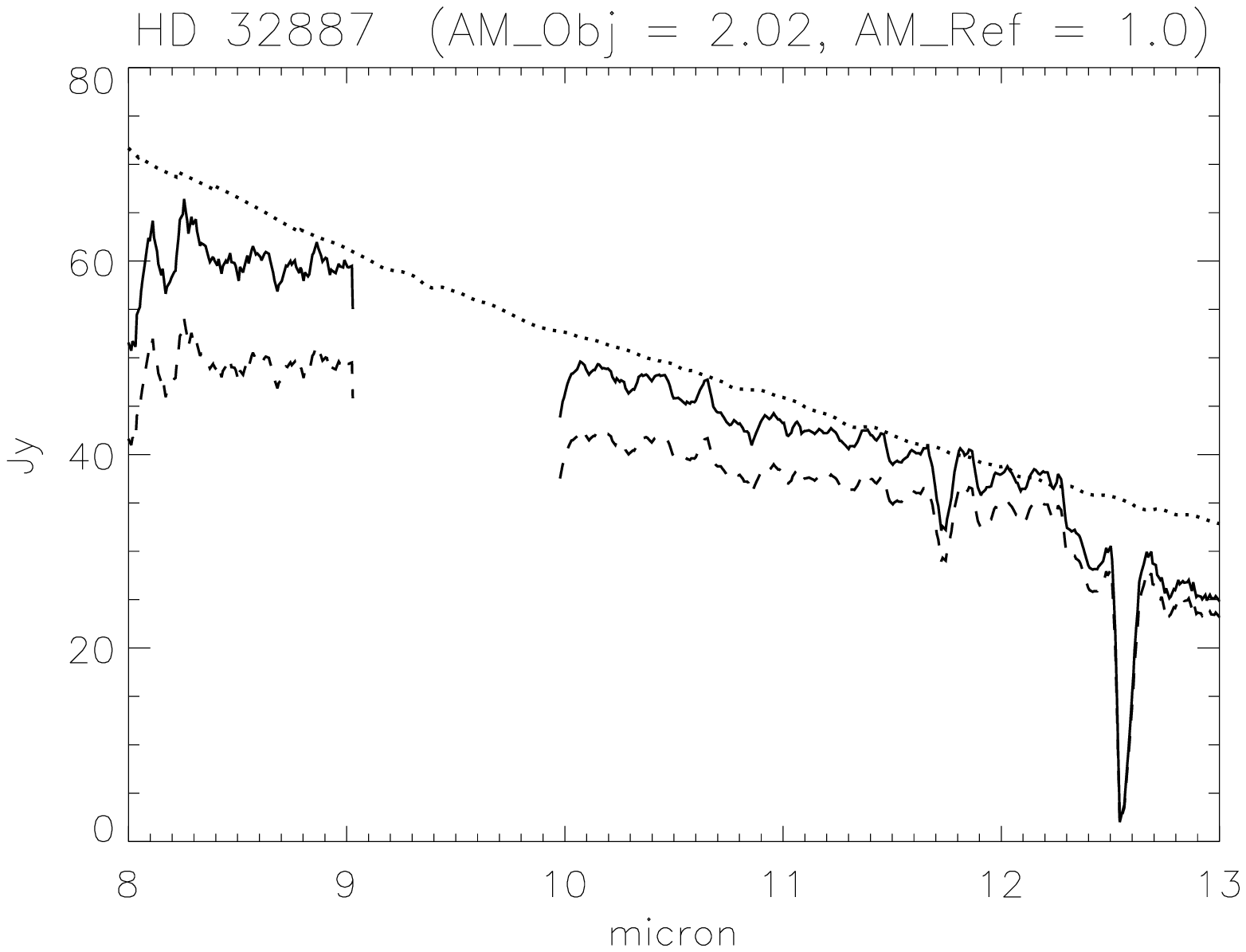}
  \end{center}
  \caption{Illustrations of our airmass correction with a calibrator
           at airmass AM = 1.0 and targets up to AM = 2.0. The target
           spectrum is shown without airmass correction ({\sl dashed
           profile}) and corrected ({\sl solid curve}). A literature
           profile is overplotted ({\sl dotted}). The quality of this 
           correction may be further improved, when we can apply
           correction factors from the other N-band filters}
  \bigskip
  \bigskip
\end{figure}

Spectra shown in Fig.\,2 were obtained in December 2002 when the
detector had a dead column between approximately 9.0\,\umu m and
9.8\,\umu m. By intention we did not correct spectral line features in
order to show how these may develop with increasing airmass between 
target and calibrator. Look especially to the CO$_2$ absorption features
at 11.73\,\umu m and 12.55\,\umu m. 

Further we derive spectrophotometric fluxes for the N11.9 filter and 
compare these with literature values. As shown in Table~2 and~3 our 
airmass correction improves the flux calibration both for targets 
observed at low and high airmass and reaches an accuracy up to 2\% within 
the literature flux.

\begin{table}[ht]
  \caption{Test of the airmass correction (AMC) with data taken on
           December 15th, 2002. The spectrophotometry in the N11.9
           filter is shown with and without our airmass correction 
           (AMC) for target and calibrator stars at varying airmass}
  \begin{center}
  \setlength\tabcolsep{4pt}
  \begin{tabular}{lcclccc}
  \hline\noalign{\smallskip}
  Object      &  AM\,(obj)   &  F\,(11.9\,\umu m) & 
  Calibrator  &  AM\,(cal)   &  no AMC            &  with AMC  \\
              &              &  [Jy]              &
              &              &  [Jy]              &  [Jy]      \\
  \noalign{\smallskip}
  \hline
  \noalign{\smallskip}
  HD 32887    &  1.03        &  41.50             &  
  HD 81797    &  1.51        &  45.58             &  43.14     \\
              &              &                    &
  HD 32887    &  1.86        &  47.82             &  43.59     \\
              &              &                    &
  HD 32887    &  2.02        &  48.46             &  43.43     \\
  \noalign{\smallskip}
  \hline
  \noalign{\smallskip}
  HD 32887    &  2.02        &  41.50             &  
  HD 32887    &  1.03        &  35.67             &  39.79     \\
              &              &                    &
  HD 81797    &  1.44        &  37.63             &  40.09     \\
              &              &                    &
  HD 81797    &  1.51        &  39.09             &  41.27     \\
              &              &                    &
  HD 32887    &  1.86        &  41.02             &  41.71     \\
  \hline
  \end{tabular}
  \end{center}
  \bigskip
  \smallskip
\end{table}

\begin{table}[hb]
  \caption{Test of the airmass correction (AMC) with data obtained 
           on Sept.\,13th, 2003}
  \begin{center}
  \setlength\tabcolsep{4pt}
  \begin{tabular}{lcclccc}
  \hline\noalign{\smallskip}
  Object      &  AM\,(obj)   &  F\,(11.9\,\umu m) & 
  Calibrator  &  AM\,(cal)   &  no AMC            &  with AMC  \\
              &              &  [Jy]              &
              &              &  [Jy]              &  [Jy]      \\
  \noalign{\smallskip}
  \hline
  \noalign{\smallskip}
  HD 187642   &  1.27        &  24.28             &  
  HD 4128     &  1.52        &  25.03             &  24.35     \\
              &              &                    &
  HD 169916   &  1.85        &  26.26             &  24.76     \\
  \noalign{\smallskip}
  \hline
  \noalign{\smallskip}
  HD 169916   &  1.85        &  22.35             & 
  HD 187642   &  1.27        &  20.64             &  21.89     \\
              &              &                    &
  HD 196171   &  1.28        &  20.74             &  21.96     \\
              &              &                    &
  HD 4128     &  1.52        &  21.31             &  21.98     \\
  \hline
  \end{tabular}
  \end{center}
\end{table}

%

\end{document}